\documentclass[amsmath,amssymb,twocolumn,amsfonts]{revtex4}
\usepackage{graphicx}
\def \beq {\begin{equation}}
\def \eeq {\end{equation}}
\def \tr {\rm Tr}

\begin{document}
\title{Comment on "Spin-selective reactions of radical pairs act as quantum measurements" (Chemical Physics Letters 488 (2010) 90-93)
}
\author{Iannis K. Kominis}

\affiliation{Department of Physics, University of Crete, Heraklion
71103, Greece}

\maketitle
Spin-selective radical-ion-pair reactions are at the core of spin chemistry \cite{steiner}. Jones \& Hore
recently introduced a master equation, which the authors claim to follow from quantum measurement theory considerations \cite{JH} and which is supposed to describe the evolution of the spin density matrix of radical-ion pairs. We are going to show that the basic assumptions of the Jones-Hore theory lead to ambiguous conclusions and hence the theory lacks self-consistency. To that end we consider for simplicity a radical-ion pair having just one recombination channel and no magnetic interactions, i.e. we set $k_{T}=0$ and ${\cal H}=0$, where ${\cal H}$ the magnetic interactions Hamiltonian.  In this case the Jones-Hore master equation \cite{JH} reads
\beq
{{d\rho}\over {dt}}=-k_{S}(\rho-Q_{T}\rho Q_{T})\label{JH}
\eeq
This master equation was derived \cite{JH} according to the following philosophy. Assume the density matrix of a radical-ion pair is initially $\rho_{0}$ and at time $t$ it is $\rho_{t}$. There are three different scenarios for what could happen within the following time interval $dt$: (i) with probability $p_{0}=1-k_{S}dt$ nothing happens, keeping $\rho_{t+dt}=\rho_{t}$, (ii) with probability $p_{S}=k_{S}dt\tr\{\rho Q_{S}\}$ a singlet recombination takes place, hence $\rho_{t+dt}=0$, and (iii) with probability $p_{T}=k_{S}dt\tr\{\rho Q_{T}\}$ a triplet projection takes place, making $\rho_{t+dt}=\rho_{T}\equiv Q_{T}\rho_{t}Q_{T}/\tr\{Q_{T}\rho_{t}Q_{T}\}$. Clearly, $p_{0}+p_{S}+p_{T}=1$ and setting $\rho_{t+dt}=p_{0}\rho_{t}+p_{S}0+p_{T}\rho_{T}$ leads to the master equation \eqref{JH}.
In other words, the Jones-Hore theory rests on the assumption, {\it built into the theory by hand}, that (in this example) the state of unrecombined radical-ion pairs acquires with time an ever more triplet character (as exemplified in the following). 

We will ask the question, what is the time evolution of the density matrix, call it $\rho_{\rm nr}$, describing unrecombined radical-ion pairs? As opposed to \eqref{JH}, which is a trace-decaying master equation, the equation for $d\rho_{\rm nr}/dt$ must be trace-preserving. It turns out that there are two ways to arrive at such an equation for $d\rho_{\rm nr}/dt$, leading to two different results.
One obvious way is to normalize $\rho$ of \eqref{JH} with $\tr\{\rho\}$, i.e. we define the density matrix $\rho_{\rm nr}$ of unrecombined radical-ion pairs as $\rho_{\rm nr}=\rho/\tr\{\rho\}$. Making use of \eqref{JH}, it easily follows that 
\beq
{{d\rho_{\rm nr}}\over {dt}}=-k_{S}\tr\{Q_{T}\rho_{\rm nr}Q_{T}\}\Big[\rho_{\rm nr}-{{Q_{T}\rho_{\rm nr} Q_{T}}\over {\tr\{Q_{T}\rho_{\rm nr} Q_{T}\}}}\Big]\label{nr1}
\eeq
Clearly \eqref{nr1} is trace-preserving, i.e. $\tr\{\rho_{\rm nr}\}=1$ for all times. 

There is yet another way to derive $d\rho_{\rm nr}/dt$. From the previous arguments, it is clear that a radical-ion pair having not recombined up to time $t$ has either remained at the state $\rho_{0}$ or has been projected to the state $\rho_{T}=Q_{T}\rho_{0}Q_{T}/\tr\{Q_{T}\rho_{0}Q_{T}\}$ at some random time before $t$. Let $w_{0}$ and $w_{T}$ denote the corresponding probabilities. Clearly, 
$w_{0}=(1-k_{S}dt)^{n}$, where $n=t/dt$, hence $w_{0}=e^{-k_{S}t}$ and $w_{T}=1-e^{-k_{S}t}$. From $\rho_{\rm nr}=w_{0}\rho_{0}+w_{T}\rho_{T}$ it then follows 
\beq
{{d\rho_{\rm nr}}\over {dt}}=-k_{S}\Big[\rho_{\rm nr}-{{Q_{T}\rho_{\rm nr} Q_{T}}\over {\tr\{Q_{T}\rho_{\rm nr} Q_{T}\}}}\Big]\label{nr2}
\eeq
Again, \eqref{nr2} is clearly trace-preserving. But then, what is the density matrix evolution of unrecombined radical-ion pairs, equation \eqref{nr1} or equation \eqref{nr2} ? Clearly, both reach the same final state, but with a different rate constant, since at $t=0$, $\rho=\rho_{nr}=\rho_{0}$ which in general can differ from the triplet state, hence $k_{S}\tr\{Q_{T}\rho_{\rm nr}Q_{T}\}_{t=0}\leq k_{S}$.

This ambiguity arises for the following reason: as explained in \cite{entropy}, the Jones-Hore  interpretation of the quantum measurement going on in radical-ion pairs makes the following two associations: (a) "measurement result of $Q_{S}$ is 0=projection to $\rho_{T}$" and (b) "measurement result of $Q_{S}$ is 1=singlet recombination". The former association is correct, but the latter is not. The truth of the matter is that the unrecombined molecules at time $t$ are comprised of (i) those that have remained in the state $\rho_{0}$, (ii) those that have been projected to $\rho_{T}$ {\bf and} (iii) those that have been projected to $\rho_{S}=Q_{S}\rho_{0} Q_{S}/\tr\{Q_{S}\rho_{0} Q_{S}\}$ at some time $t'$ prior to $t$ {\it but until $t$ have not recombined}. However, possibility (iii) is missing from the Jones-Hore theory applied to this example.

We elaborate a bit further.  No matter what the final theory will be, it is clear that the change $d\rho$ that updates $\rho_{t}$ into $\rho_{t+dt}$ can be written as $d\rho=-d\rho_{\rm r}+d\rho_{\rm nr}$, where the first term is the change in $\rho_{t}$ due to recombined radical pairs, and the second (traceless term) the state change of the ones that did not recombine within $dt$. Accordingly, $\tr\{\rho\}$ will change like $\tr\{\rho_{t+dt}\}= \tr\{\rho_{t}\}- \tr\{d\rho_{\rm r}\}$. Suppose that between $t=0$ and $t=dt$ a fraction $x$ (proportional to $dt$) of radical pairs initially in the state $\rho_{0}$ recombine. Then $\rho_{dt}=\rho_{0}-x\rho_{0}+d\rho_{\rm nr}$ and $\tr\{\rho_{dt}\}=1-x$, thus to first order in $dt$ it follows that
\beq
d\Big({{\rho_{t}}\over {\tr\{\rho_{t}\}}}\Big)_{t=0}={{\rho_{0}-x\rho_{0}+d\rho_{\rm nr}}\over {1-x}}-\rho_{0}\approx d\rho_{nr}
\eeq
So indeed, at least for early times, $\rho/\tr\{\rho\}$ accounts for the state of unrecombined molecules. Thus the rate of change $d\rho_{\rm nr}/dt$ calculated from \eqref{nr1} {\it must} coincide with the one calculated from \eqref{nr2}. This is not the case with the Jones-Hore theory.

The author would like to acknowledge several stimulating and helpful discussions with Prof. Ulrich Steiner.

\end{document}